\documentclass[a4paper,dvips,12pt]{article}
\usepackage{axodraw}
\usepackage{array,tabularx,epsfig,mathrsfs,rotating}
\usepackage{graphicx}
\usepackage{ifthen}
\usepackage{fancybox}

\usepackage{amsfonts}

\usepackage{caption}
\newcommand{\beq}{\begin{equation}}
\newcommand{\eeq}{\end{equation}}

\newcommand{\Ec}{\mathcal{E}}

\begin{document}
\vspace*{+20mm}
%\begin{flushright}
%Preprint SINP MSU 05--*** \\
%March-23/August-08\,, 2006, v4.6\\
%\end{flushright}  
\begin{center}
{\large \bf On Field Emission  in High Energy  Colliders Initiated  \\
by a Relativistic Positively Charged Bunch of Particles}\footnote{Preliminary
results were presented at ICHEP'06, Moscow, Russia, 27.07-02.08 2006.}

\vspace{5mm} 
 {\large B. B. Levchenko}\footnote{Electronic address: levtchen@mail.desy.de  }\\
\vspace*{2mm}
{\it Skobeltsyn Institute of Nuclear Physics, Moscow State University \\
 119992 Moscow, Russian Federation}\\

\end{center}

\abstract {\small 
\vspace*{-7mm}
\noindent\begin{quote}
The design of the LHC and future colliders  aims their  operation
with high intensity beams, with   bunch population, $N_p$, of the order 
of $10^{11}$. This is dictated by a desire to  study  very 
rare processes with maximum  data sample. HEP colliders  are  engineering 
structures  of many kilometers in length, whose  transverse  compactness is 
achieved by the application of the superconducting technologies and 
 limitations of  cost. 
 However the compactness of the structural 
elements conceals and potential danger for the stable work of the accelerator. 
This is because a high intensity beam of  positively 
charged particles (protons, positrons, ions) creates around  itself an 
electric self-field of  very high intensity, $10^5 - 10^6$  V/cm. 
Being located near the conducting surfaces, at the distances of 1-20 mm 
away from them, the field of such bunches activates the field emission of 
electrons from the surface. These electrons, in addition to electrons from 
the ionization of residual gases, secondary electrons and electrons knocked 
out by synchrotron radiation, contribute to the development of a dense 
electron cloud in the transport line. The particles of the  bunch, being 
scattered on the dense  electron cloud  with  $N_e\sim N_p$,  leaves the 
beam and  may cause noticeable damage. The paper presents an  analysis of 
the conditions, under which the  field emission in the LHC collimator system  
may become a serious problem. The analogous analysis of a prototype
of the International Linear Collider (ILC)   project, USLC, 
reveals that a noticeable field emission will 
accompany positron bunches on their entire path during acceleration.
 
\end{quote}
}
\vspace{1.0cm}

\newpage

\section{\large Introduction}

Searches for rare processes and the production of new particles
require collider experiments  to be runs
 with the highest possible luminosity, given by the standard expression
\beq
\mathcal{L}\,\sim\,\frac{n_b\cdot N_1\cdot N_2}{\sigma_x \sigma_y}.
\label{I.1}
\eeq
To achieve this goal it is necessary to have beams with  maximal bunch population,
$N_i$, large number of bunches, $n_b$, and bunches with minimal transverse sizes,
$\sigma_x$, $\sigma_y$.
 The electric and magnet self-fields created 
by an intense particle bunch,
with modification by the accelerator  environment (beam
pipe, accelerator gaps, magnets, collimators, etc.) due to induced surface 
charges or currents, can have  very high values  at a distance of the order
of a few millimeters from the bunch.

It is a well known fact that under  the influence of an  electric field  $F$ 
 of the order of 10$^5$  V/cm at room temperature the smooth surface of solids  
(conductors, semiconductors, dielectrics) and liquids  emits 
electrons\footnote{We denote  the electric field by $F$ follow the established 
tradition  in   publications  devoted to  field  emission.}
 (see  \cite{litt} and references  therein). 
The phenomenon, known now  as electron field emission (FE), was discovered by Robert 
Wood \cite{wood} in 1897.
Later, in 1928, Fowler and Nordheim \cite{FN}  proposed a quantum  theory 
of field emission from metals  in terms of electrons  tunneling through a potential barrier.
Application of a high electric field to the metal  produces a triangular shaped potential
energy barrier through which electrons, arriving at the metal surface, may quantum 
mechanically tunnel. 

Such electron emission usually precedes electrical breakdown in vacuum and is
called pre-breakdown emission.
We note that the maximum field strength that can be maintained between two
conductors in air is limited to less than about 10$^4$ V/cm, above which
dielectric breakthrough leads to the formation of a plasma.
With semiconductors, of the order of 10$^6$ V/cm can be maintained.
Fields of the order of 10$^8$ V/cm = 1 V/\AA \ can also be established
within 10$^3$ \AA \ of a metal tip with a tip radius of less than 10$^3$ \AA, 
provided breakthrough is avoided by working in ultrahigh vacuum.
In  fields larger than  1 V/\AA \ a variety of 
 processes start to develop: 
 field desorption\footnote{ Field desorption
 is a process in which an atom or molecule leaves a solid surface due
to the influence of a high electric field. The atom or molecule usually departs 
from the surface in a charged state as an ion. }, 
field evaporation\footnote{Field evaporation
 is the removal of lattice atoms as singly or multiply charged
 ions from a metal in a strong electric field F of the order of several 
V/\AA, as  occurs at field ion tips.},
vacuum discharge.

In this paper,  we discuss the field emission phenomena activated 
in strong fields created by a positively charged particle beam.
Primary attention is paid to  the field emission processes and
beam induced multipacting in the LHC collimator system. The  primary 
and secondary collimators in the IR3, IR6 and IR7 insertions are  made of graphite 
\cite{lhc-c0}-\cite{lhc-c4} 
with parallel jaw surfaces and small  gaps, $\sim$2 mm, at 7 TeV.
As shown below, for the LHC bunch parameters  and 
structural  features of the LHC collimator system, the density of the electron 
emission current  can reach a significant magnitude,
which in turn can lead to a "thrombosis"  of the collimators.

The paper is organized as follows. In Section 2 we briefly review  the 
classical Fowler-Nordheim  theory for point cathodes, following which we give 
an estimation 
of the tunneling time and discuss deviations from the F-N theory due to extensions   
to non-zero temperatures and "real" broad-area cathodes.
In Section 3 we discuss electrical self-fields generated by charged particles 
encountered
in  high energy physics: an elementary particle, a relativistic dipole,
a particle bunch.
In Section 4, the results of studies made with the use of the modified F-N theory
are presented. Here we discuss the field emission  in the LHC
collimator system and the main linac beam pipe of the US linear collider. 
Finally, Section 5 discusses the  dynamics of emitted electrons,
the electron multiplication chain, and the development of an electron cloud
 in the self-fields of a bunch.

\section{\large Field Emission in Strong Electric Fields}
%---------------------------------------------------------------------

\subsection{Fowler-Nordheim theory}
%------------------------------------------

In the framework of  Fowler-Nordheim (F-N) theory,  the current density of field 
emission of electrons from a metal can be written in the following form 
\cite{FN},\cite{SB}-\cite{Mod}
\beq
J_{FN}\,=\,e\int n(\Ec_x )D(\Ec_x,F)d\Ec_x,
\label{3.1}
\eeq 
where $D(\Ec_x,F)$ is the penetration coefficient and $n(\Ec_x )$ is the number
of electrons at the energy $\Ec_x$ incident in the x-direction on the surface barrier 
from inside of the metal.

An electron outside a metal is attracted to the metal  as a result of the 
charge it induces on the surface (image force).
In the externally applied accelerating electric field $F$, the potential energy of the
electron is\footnote{In the present paper we  adopt the Gaussian CGS system.}
\beq
V(x)\,=\,-\frac{e^2}{4x}\,-\,eFx,\ \ \ \ {\rm when} \ \ \ x>0,
\label{3.2}
\eeq 
where  $x$ denotes the distance from the surface
and the first term accounts for the image potential. With use of 
the potential energy (\ref{3.2}) and the Fermi energy distribution of electrons 
in the conduction band,  one finds  
\cite{FN},\cite{SB}-\cite{Mod} that
\beq
J_{_{FN}}(F)\,=\, A \frac{F^2}{\varphi\cdot t^2 (y)}\exp \Big\{-B\frac{\ 
\varphi ^{3/2}}{F}\vartheta (y) \Big\}\, ,
\label{eq:n1}
\eeq 
where $J$ is the current density  in A/cm$^2$, $F$ is electric field  on the  surface 
in V/cm, and  $\varphi$ is the work function in eV.  The field-independent constants A and  B 
 and the variable  $y$ are 
\beq
A=\frac{e^3}{8\pi h}\,=\, 1.5414\cdot 10^{-6}, \ \   B=\frac{8\pi
\sqrt{2m}}{3eh}\,=\,6.8309\cdot 10^7,
\ \  y= \frac{\sqrt{e^3 F}}{\varphi} = 3.7947\cdot 10^{-4} 
\frac{\sqrt{F}}{\varphi}
\label{eq:n2*}
\eeq 
where $-e$ is the charge on the electron, $m$ is the electron mass
and $h$ is  Planck's constant. The numerical values of $A$ and $B$  correspond 
to recent values of the physical constants   \cite{PDG}.  
We note that under field emission conditions, 0$<y \le$1.

The Nordheim function $\vartheta (y)$ takes into account  a lowering of the potential 
barrier due to the image potential (the Schottky effect) and  its distinction from an 
idealized triangular shape. 
The function $t(y)$ in the denominator of equation (\ref{eq:n1}) is defined as
\beq
t(y)\,=\,\vartheta (y)-(2y/3)(d\vartheta /dy).
\label{eq:n5}
\eeq  
The function $\vartheta (y)$ varies from $\vartheta (0)=1$ to $\vartheta(1)=0$
 with the increase in field strength, however  $t(y)$  is quite close to unity 
at all values of $y$. 

For  a typical metallic $\varphi$ of 4.5 eV, fields of the order of 10$^7$ V/cm are needed to
have measurable emission currents. In considering magnitudes, one must always keep in mind
the rapid variation of the exponential function. For instance, an increase in $F$ of only a
factor of two from $1\times 10^7$ to $2\times 10^7$ V/cm increases the current density by
15 orders of magnitude (from $10^{-18}$ to $10^{-3}$A/cm$^2$) ! 

At a field strength of the order of
$F_{cr}=\varphi^2/e^3=6.945\cdot 10^{6}\cdot \varphi^2$ V/cm 
the height of the potential barrier  vanishes  and $\vartheta (1)=0$. 
For instance, for copper $\varphi_{Cu}=4.65$ eV giving 
$F_{cr}(Cu)=1.5\cdot 10^{8}$ V/cm,
and  similarly for graphite, $\varphi_{gr}=4.6$ eV,  
$F_{cr}(gr)=1.47\cdot 10^{8}$ V/cm.
At this field level one would expect the
orderly bound states characteristic of the solid to lose their integrity.

For a long time only tabulated values of $\vartheta (y)$ and $t(y)$
\cite{BKH}  were used in calculations, see \cite{MG}-\cite{Mod}. 
Recently \cite{bbl1},  several  parameterizations of the functions  
$\vartheta (y)$  and $t(y)$  were proposed.

The theory of field emission from metals has been subjected to fairly extensive
verification. A variety of methods have been employed over many
years for the experimental measurements of the emission current as a function
of  the field strength, the  work function and the energy distribution of  
the emitted electrons \cite{GM}-\cite{Mod}.
The F-N theory (\ref{eq:n1}) of electron emission from plane
and uniform metal surfaces (single-crystal plane) at $T\approx 0$
may therefore be considered well established on 
experimental basis as well as on theoretical grounds.

%We have to stress that the formula (\ref{eq:n1}) apply only to emission from plane
%and uniform metal surfaces (single-crystal plane) at $T\approx 0$.

\subsection{Tunneling Time}

Using the Heisenberg uncertainty principle one can  estimate the tunneling time.
For electrons near the Fermi level, the uncertainty  in their momenta, $\Delta p$, should
correspond a barrier of height $\varphi$ where

\beq
\varphi\,=\,\frac{(\Delta p)^2}{2m},\ \ \ \ \Delta p\,=\,\sqrt{2m\varphi}.
\eeq
If the corresponding uncertainty in position,
\beq
\Delta x \simeq \frac{\hbar}{2\sqrt{2m\varphi}}
\eeq
is of the order the barrier width
\beq
\Delta x \sim \frac{\varphi}{eF},
\eeq
one expects to find electrons being emitted.

On the other hand, the uncertainty in energy is of the order
\beq
\Delta \epsilon \approx eF\Delta x
\eeq
and therefore the estimated value of the tunneling time is
\beq
\Delta t \approx \frac{\sqrt{2m\varphi}}{eF} 
\approx 3.37\cdot 10^{-8}\frac{\sqrt{\varphi}}{F}.
\eeq
Here the field strength $F$ is in V/cm,  the work function $\varphi$
is in eV and the tunneling time $\Delta t$ is given in seconds. As an example, 
for graphite and  a field of  $10^{6}$ V/cm 
this gives $\Delta t = 7.23\cdot 10^{-14}$ s, quite a short time scale,
when the field is large.
If the field is created by a relativistic bunch of  length $L$, the field is
acting on a certain area of the surface during the time $\tau = L/c$. For 
a LHC bunch of $L$=7.55 cm one obtains $\tau = 2.52\cdot 10^{-10}$ s, which is long
in  comparison with  the emission time $\Delta t$.

%Work functions of graphite
%5.0 eV (D.S.Y. Hsu, Appl. Phys.Lett.80(2002)2988 )
%4.6 eV (Gang Zhou, Yoshiyuki Kawazoe, Chem. Phys. Lett. 350 (2001)386 )
%4.0 eV (Fomenko V.S., Handbook of Thermionic Properties, (1966) Plenum Press
%Data Division, NY (http://www.netl.doe.gov/publications/proceedings/05/UBC/
%pdf/Cangialosi_characteristics_paper.pdf
%4.7 eV Liu Xinghui et al, Physics B344(2004)243

\subsection{Deviations from the Fowler-Nordheim theory}
%--------------------------------------------------------------------
\subsubsection{Temperature dependence}

The main equation (\ref{eq:n1}) of the  F-N theory was derived for an idealized
metal in the framework of the Sommerfeld model, with an ideally flat  surface and  
at a very low temperature, $T\approx 0$.
The temperature dependence of the field emission current (FEC) is completely 
connected with the change of the spectrum  of electrons in the metal with an increase in $T$.
Therefore, at non-zero temperatures the F-N theory must be modified to take into account
the thermal excitation of electrons above the Fermi level.
For the so-called  extended field emission region,  Murphy and  Good \cite{MG}
 (see also \cite{Chr1},\cite{Mod})  obtained the following elegant equation
\beq
J_{_{FN}}(F,T)\,=\,\frac{\pi \omega}{\sin \pi \omega}J_{_{FN}}(F,0),
\label{s2.3.1}
\eeq 
which account for  the temperature dependence of the FEC.
 Here $\omega = k_{_B}T/k_{_B}T_{_0}$ and 
\beq
k_{_B}T_{_0}\,=\, \frac{2}{3}\frac{F}{B t(y)\sqrt{\varphi} },
\label{s2.to}
\eeq 
where $k_{_B}$ is the Boltzmann's constant, and $T$ is the absolute temperature in K. 
It can be shown \cite{MG} that equation (\ref{s2.3.1}) is a valid approximation when the
following two conditions are satisfied:
\beq
\omega <  \Big [1+\frac{1}{\Gamma_1}\Big ]^{-1}, \ \ \ \Gamma_1 =\frac{\varphi (1-y)}{k_{_B}T_{_0}}
-\frac{2}{\pi}\Big (\frac{2}{y}\Big )^{1/2}t(y)
\label{s2.g1}
\eeq 
and
\beq
\omega  < \Big [1+\frac{1}{\Gamma_2}\Big ]^{-1}, \ \ \ \Gamma_2 \simeq
\Big (\frac{2\varphi }{k_{_B}T_{_0} t(y)}\Big )^{1/2}.
\label{s2.g2}
\eeq 
At very low temperatures, when $\pi\omega$ is small, equation  (\ref{s2.3.1}) reduces 
to (\ref{eq:n1}).
By expanding $\sin \pi \omega$ in a series, one gets  for  practical use the
formula
\beq
 J_{_{FN}}(F,T)/J_{_{FN}}(F,0)=1+1.40\cdot10^8(\varphi/F^2)T^2.
\label{s2.3.2}
\eeq 
It is easy to estimate using  (\ref{s2.3.2}) that for $\varphi=$4.5~eV  
at room temperature, $T=300$K, and
$F=1\times 10^{7}$ V/cm and $2\times 10^{7}$ V/cm,
 the temperature factor in  (\ref{s2.3.1})  equals 1.57 and 1.14, respectively.
Thus, the temperature factor appear to be a sizeable correction.

The energy distribution of the emitted electrons is determined by two effects: the
low-energy slope by the tunneling probability, and the high-energy slope by the electron
distribution in the metal, and thus by the temperature $T$ of the emitter. For increasing
fields, the width of the energy distribution will grow approximately with the field, while
the position of the spectrum stays close to the Fermi level, $E_{_F}$  \cite{Mod}:
\beq
\frac{dJ}{d\epsilon}\,=\, \frac{J_{_{FN}}(F,0)}{e\,k_{_B}T_{_0}}\cdot
\frac{\exp (\epsilon/k_{_B}T_{_0})}{1+\exp(\epsilon/k_{_B}T)}.
\label{s2.ed}
\eeq 
Here $\epsilon = E-E_{_F}$ and $E$ is the kinetic energy of the electron.

\subsubsection{ Anomalous High FEC  }
%-----------------------------------------------

Since the first experimental verifications of the F-N field emission theory, it was
noticed that for broad-area cathodes the electrical breakdown typically occurs for field
values one or two orders of magnitude smaller and pre-breakdown FEC is  many orders of
magnitude greater than the values predicted by the F-N equation.  Nevertheless the observed
emission current follows the F-N law, provided one makes the substitution
$F\rightarrow F_{\!_{eff}}=\beta_{_{FN}}F$ for all occurrences of $F$ in (\ref{s2.3.1})  
\cite{lewis,slivkov,STW,noer1}. 
Here $\beta_{_{FN}}$ is known as the field enhancement factor.
Generally,
$\beta_{_{FN}}$ values in the range 50 $<\beta_{_{FN}}<1000$ have been observed in many
experiments.
Emission does not occur homogeneously over the surface but is rather concentrated in $\mu$m
and sub-$\mu$m sized spots \cite{litt,slivkov,xu}.  Then electrons
are concentrated  in tiny jets and with further increase in voltage, this ultimately leads
to breakdown.
Physics of the local field enhancement and a vacuum arc discharge, in general, is very complex
and still not completely understood. For reviews see  
\cite{slivkov},\cite{laffetry}, \cite{LMP}. 
As one sees from  (\ref{eq:n1}) ,  FEC increases both with an increase of $F$ as
well  as with decrease of $\varphi$, or due to a combination of these two factors. 
There are a number of mechanisms which lead to an increase of FEC.
Here we enumerate  some of them.

%The maximum field strength at the apex is
%\beq
%F_0\,=\, U\Big ( \frac{\alpha}{r}+\frac{2(1-\alpha)}{r\ln\frac{4R}{r}}\Big )
%\eeq
%where U is anode potential, R is the distance tip-anode, r is the radius of
%curvature in the apex, $R\gg r$.

{\bf Geometrical field enhancement.} The real surface of solids is not perfect.
For many years, the anomalous emission was universally explained by invoking a
'projection model' \cite{lewis,slivkov,noer1,noer2}. This model assumes the presence 
on the cathode 
surface of a number of microscopic projections (crystalline defects, impurities, whiskers 
or dust 
particles), sharp enough to cause a geometric enhancement of
the local field at the projection tip to a value some 100 times greater than the
nominally applied field. The projection model  was confirmed by 
experimental evidence \cite{litt,slivkov,noer2}  obtained with 
 shadow/scanning electron microscopy. 
Even on optically polished   cathodes made of stainless steel, tungsten, copper or aluminum,
 needle-like projections about 2 $\mu$m high have been found.
These  projections are capable of producing field enhancements of the order of 100  
 at pre-breakdown emission sites.

{\bf Resonant tunneling.} To account for enhanced emission from gas condensation,
another model assumes that ad-atoms are responsible for creating localized energy
levels near the surface. One dimensional calculations \cite{duke} show that the tunneling
process of electrons with energies close to the localized  states can be resonantly
enhanced. The calculations predict that tunneling is enhanced by up to a factor of
10$^4$ for adsorbates less than a single monolayer thick.
For example, adsorbed water with its strong dipole moment can be a critical 
factor in enhanced  electron field emission \cite{halb}. 
Water  is certainly one of the main  adsorbates on
the conductor surface, especially if the surface is not baked. There is also 
observation that adsorbed oxygen enhances field emission \cite{shu}.

{\bf Surface states (SS).}
By the level of  complexity, this is a completely different mechanism  
resulting  the  FEC increase due to a lower  electron work function of SS.
A semi-infinite crystal extending from $x=-\infty$ to $x=0$  can be   viewed as a stack of 
atomic layers parallel to the given crystallographic plane but, in contrast to the infinite 
crystal, all  layers cannot  in general be  identical.
% Obviously, the layers which lie away from the surface, deep into the solid, will be unaffected 
% by the presence of the surface, but the top few layers are bound to be different from 
% the bulk layers in more than one way. In the first instance the relative position of 
% the atomic nuclei at the surface may be different from those in the bulk. For metals 
% where a layer consist of a single plane of atoms this difference usually consists of a
% simple dilation away from or contraction towards the bulk of the metal 
% of the top one or two layers. 
Consequently, the potential the electron sees  at the  metal-vacuum interface 
is unavoidably different from that in the bulk material because the electronic 
charge distribution  is different  on the surface. 
On metal surfaces, the density of electrons is high, of the order of 10$^{15}$ cm$^{-2}$. 
That gives rise to  creation and decay of a variety 
of  electronically excited SS \cite{AL,MP,Mod}.  
The term applies because the wave function of such a state is localized 
near the surface, decaying "exponentially"  on either side of it.
The surface states are classified  as Tamm \cite{tamm} and Shockley 
\cite{ws39} surface states (TSSS),  and image potential states (IPS) \cite{echen}.
Tamm states  are more localized and arise when the potential 
in the top surface layer  is significantly distorted, whereas Shockley states show
a strong free-electron-like dispersion and may be interpreted as dangling bonds of surface
atoms.  Such SS are described by wave functions whose center of gravity lies in
the immediate neighborhood of the metal-vacuum interface on the metal side of it.
Self-consistent  calculations \cite{gay} and angle-resolved photoemission measurements
\cite{heimann}, have shown that such states exist on the (100) plane of copper.
Employment of  scanning tunneling microscopy at low temperatures allows one to detect the 
surface topology and view SS of many noble metals. For a state-of-the-art review 
see \cite{hoefer3}.

Image potential states  \cite{echen} (see also \cite{SH}-\cite{hoefer3}, \cite{AL}),
%are  the quantized excited states of electrons that exist in front of many metal surfaces  
arise for an additional electron in front of the surface. The screening of the charge by
the metal electrons give rise to a long-range  Coulomb-like attractive image potential which leads 
to  bound states which form a  Rydberg series with energies $E_n$, where
\beq
E_n\,=\,\frac{-0.85}{(n+a)^2}\, eV,\ \ \ \ n=1,2,...,\ \ \ \ 0\leq a \leq 0.5
\eeq
converging toward the vacuum energy. Multiphoton photoemission spectroscopy 
allows  us to identify IPS on graphite \cite{lehm}.
For IPS on graphite a quantum defect of 
$a = -0.04\pm0.05$ is obtained. This demonstrates that IPS on graphite exhibit an almost
ideal hydrogen-like behaviour.
The maximum of the IPS wave function density  with n=1 is located several
Angstroms away from the surface at the vacuum side \cite{hoefer1,zamkov1}. 

While the energy  levels of TSSS are spread around the Fermi level, 
the energy  levels of IPS are allocated very close to the vacuum level, 
$ E_n \geq  -$1.0 eV. That is, the effective work function of TSSS or IPS 
is equal to  $\varphi_{eff}\,=\,|E_n|$.

\begin{figure}[t]
\begin{center}
\includegraphics[height=10.0cm,width=12.0cm]{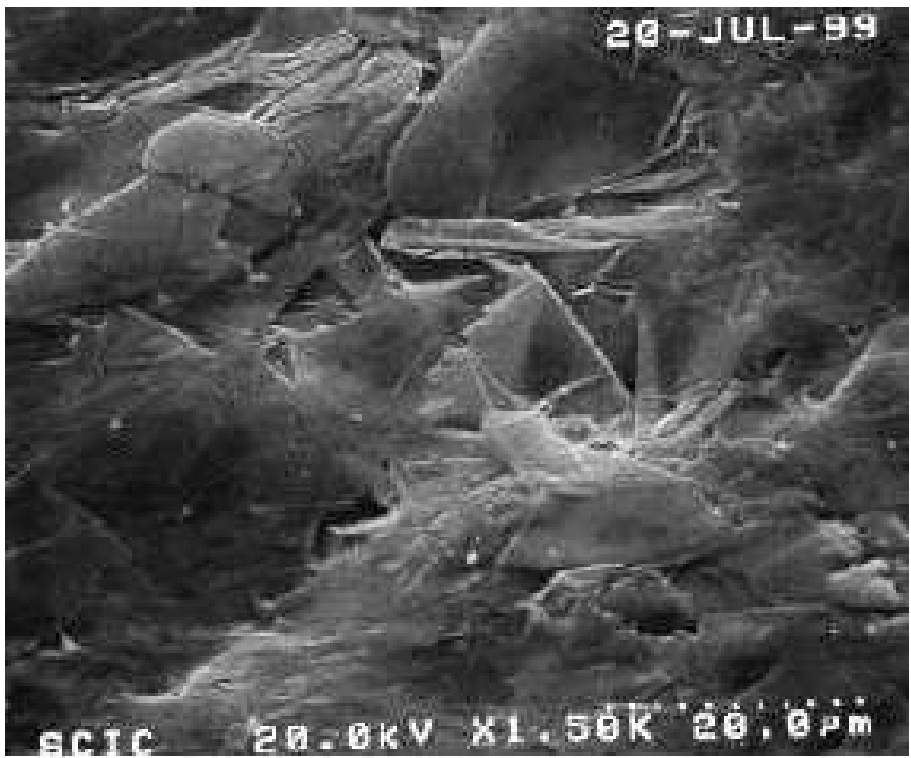}
\end{center}
\vspace*{-0mm}
\caption{ \small Typical rough graphite surface \cite{RHZL}.}
\label{fig1} 
\end{figure}

{\bf Fullerenes, nanoparticles, nanotubes etc }\cite{Ele95}-\cite{Ele02}.
The LHC collimator jaws are made of graphite \cite{lhc-c00}-\cite{lhc-c5}. 
Graphite can be polished only mechanically, and therefore may serve as a large area emission
surface. The simplest explanation for this  is that the surface morphology \cite{hunt}
consists of countless fractures, protrusions, edges, and other features which 
are readily demonstrated to emit electrons with field enhancement. Fig {\ref{fig1}}
shows the typical graphite surface  at high magnification \cite{RHZL}.
In conjunction with the previous discussion,  we shall consider briefly  
some properties of graphite and its derivatives.

Graphite consists of layers which are formed by regular hexagons.
Partial thermal decomposition of graphite layers (as a result of heating 
at higher local emission current densities) can produce on 
the surface fullerene molecules,  nanoparticles and also long tiny tubes, nanotubes. 
Under the influence of a high voltage  field these objects are able to emit electrons.        

The $C_{60}$ molecule occupies the central place among fullerenes. This molecule resembles
the commonly-used surface of a football. The carbon atoms are distributed on a spherical surface at the
vertices of twenty regular hexagons and twelve regular pentagons. The radius of the 
$C_{60}$ molecule, deduced from X-ray structure analysis \cite{kraetsch}, is 0.357 nm.
The common structural components found in  graphite and the $C_{60}$ fullerene molecule determine
the nature of the process of formation of fullerenes by the decomposition of graphite. Moderate
heating of graphite breaks the bonds between the separate layers and an evaporated layer
then splits into separate fragments. These fragments are combinations of hexagons  one of
which becams used to form a cluster. 

The process of formation of fullerenes from graphite yields further various structures which are
composed, like graphite, of six-member carbon rings. These structures are closed and empty
inside. They include nanoparticles and nanotubes. The nanoparticles are closed structures
similar to fullerenes, but of much larger size. In contrast to fullerenes, they may contain
several layers. Such multilayer spherical structures are called onions.
The study \cite{jaya} of field emission properties of diamond-like carbon films indicate that
the films have a faceted morphology with carbon cluster size in the range 150-400 nm.
The fitting of I-F (current-electric field) data to the F-N equation gives 
$\beta_{_{FN}} \simeq$160, assuming $\varphi$ = 5 eV.

Carbon nanotubes (CNT) are elongated  structures with  surfaces formed by regular hexagons.
Separate extended graphite fragments, which are then vent into a tube, form the base of 
CNT. These tubes are up to several microns long with diameters of 
a few nanometers. They are multilayer structures with rounded ends.
One end is attached to a surface and the other is free.

As follows from direct measurements \cite{heer}-\cite{Cho}, CNT have quite a high
field emission capability. This property centers around the extraordinarily small diameter 
of CNT, which provides a sizeable enhancement of the electric field strength near the
CNT cap in relation to that averaged over the entire volume of the interelectrode gap.
For instance, at a voltage near 650 V, an emitting area of about 1 mm$^2$ provides an
emission current of order 0.5 mA.
The measured emission  characteristics correlate well with the F-N equation, 
with the condition that the value of the electric field strength is taken at the point of
the electron emission. Since this point is situated close to the sharpened top of the
CNT cap, the  local magnitude of the electric field considerably exceeds its mean
value. Thus, the above mentioned effect of  field enhancement becomes involved, 
with the magnitude of  $\beta_{_{FN}}$  estimated to be of the order of 1000, 
assuming the work function to be close to $\varphi$ = 5 eV. 

 High emission properties of CNT cannot be attributed only to their high aspect ratio. As
demonstrated in model calculations and measurements \cite{zamkov1,zamkov2}, single- and 
multiwalled CNT generate both IPS as discussed above and a new class of surface 
states, the tubular image  states, owing
to their quantized centrifugal motion. Measurements of binding energies and the temporal
evolution of image state electrons were performed using femtosecond time-resolved
photoemission. A cluster of  IPS with n=1 is located near 0.75 eV below
the vacuum level. These data are in agreement with results obtained 
in the course of studies of the  emission properties of CNT \cite{zhirnov,carnah}.
The values of the effective work function calculated from 
the Fowler-Nordheim plot are in the range  0.3 $-$ 1.8 eV and the highest value 
of  $\beta_{_{FN}}$ extracted from the data
is  of the order of 1300.

\section{\large The Electric Field of Relativistic Charges }
%--------------------------------------------------------------------

\subsection{\large Single Charged Particle}
%----------------------------------------------------------------

The electric field of a charged particle at rest is spherically symmetric. If the
particle is moving with a uniform velocity ${\bf v}$ in an inertial frame ${\bf S}$,
its electric field is deformed: along the direction of motion the
electric field  becomes weaker  by a factor  $\gamma ^2$, 
while in the transverse direction the electric field is enhanced by a 
factor $\gamma$. Here, $\gamma$ denotes the  particle Lorentz factor.

Let $(x,y,z)$ be the coordinates and ${\bf \vec{r}}$ the position vector of the point 
P in ${\bf S}$,  the z-axis of the frame being taken along the direction of
motion of the charged particle. Then the components of the electric field  
produced by a rapidly moving  charge $q$ are \cite{SR}, \cite{FLS}

\beq
\vec{F}\,=\,\frac{\kappa\,q\, \gamma}{[x^2+y^2+\gamma ^2 (z-vt)^2\,]^{3/2}}\,
(x,y,z-vt),
\label{eq:1}
\eeq 
where the value of $\kappa$  depends on the system of units used,   in our case 
$\kappa=1$ \footnote{Nevertheless, we keep the symbol $\kappa$ in equations below to control 
their dimensions. }.
We use the particle charge $q$ rather than the elementary charge  $e$ to include particles
with multiple charges such as ions for which  $q=Ze$.
 At $t=0$ and $z=r\, cos\,\theta $, where $\theta$ is the angle which the vector
 ${\bf \vec{r}}$ makes with the z-axis, 
equation (\ref{eq:1}) can be rewritten in the form
\beq
 \vec{F}\,=\, \kappa\frac{q\,\gamma}{ r^2} \Big [ 
 \frac{1-\beta ^2}{1-\beta ^2\, sin^2\,\theta}\Big ]^{3/2}\,
\frac{\bf \vec{r}}{r},
\hspace*{5mm}
\vec{B}\,\sim \vec \beta \times \vec F.
\label{eq:2}
\eeq
Thus, the magnitude of the electric field at $\theta\,=\,0^{\circ}$ or 
$\theta\,=\,180^{\circ}$, $F_{\parallel}$, is  given by
\beq
 F_{\parallel}\,=\,  \kappa\frac{q}{\gamma ^2\,r^2},
\label{eq:3.3}
\eeq
while in the transverse direction, the magnitude of the electric field is
\beq
 F_{\perp}\,=\,  \kappa\frac{q\,\gamma }{r^2} .
\label{3.15}
\eeq
As  seen from Eq.(\ref{3.15}), the increase of the strength of the electric
field in the transverse direction corresponds efficiently to an increase of
the particle electric charge, $Q_{eff}=\gamma Ze$.

In accelerators, charged particles move in front of  conducting surfaces
and one has to account for the  image charge induced by  the particle. 
In this section we  discuss  only the case when a positively charged 
particle is moving parallel to a grounded conducting plane.
The particle and  the image charge form a dipole. The complete form of 
the electric  field vector of the relativistic dipole is presented in 
 Appendix . 
%The electric field just outside the conductor is normal to the plane.

Let the surface of the  conductor coincide with the (y,z)-plane and  (0,y,z) be 
 a point  on the plane.
The electric field at this point is normal to the surface and is directed into it.
The components of the field from a positive point charge, located at 
the distance $h$ above the plane, are (see Appendix)
\beq
\vec{F}\,=\,\frac{2\kappa q\gamma }{\big [ h^2+y^2+\gamma^2(z-vt)^2\big ] ^{3/2}}
(-h,0,0).
\label{eq:a}
\eeq
Thus,  at $t$=0 the magnitude of the electric field at the point  directly 
beneath the positive charge is
\beq 
F_{\perp}\,=\,\frac{2\kappa q\gamma}{h^2}.
\label{eq:b}
\eeq
We find that accounting for  the image charge  doubles the electric field strength
on the surface.

From equation (\ref{eq:b}) we  conclude that only  particles  with ultra-high energies 
are able to produce significant field strengths at macroscopic distances. 
 Let us substitute in (\ref{eq:b}) the numerical values of 
constants. That gives
\beq
 F_{\perp}\,=\,2.88\cdot 10^{-7}\frac{ Z\gamma}{h^2}\,,
\eeq
where $h$ is the transverse distance in cm, $F$ is the electric field  in V/cm. For
instance, the field strength of 1\,V/cm  at a distance  1 cm from the particle 
is generated by a positron  of energy 1.75 TeV or a proton   
of energy $3.25\cdot 10^3$ TeV. Thus, only particles  of ultra-high
energies (e.g. cosmic rays) are  capable to generate a large field strength
at macroscopic distances.

\subsection{ Bunch of Charged Particles}
%-----------------------------------------------
%Now consider  a bunch of positively charged particles. 

Let us consider a bunch of $N$ positively charged particles uniformly distributed 
within a  circular cylinder of  length $L$.  
Suppose that the bunch axis  is along the coordinate  z-axis and the bunch is moving along 
the z-axis with a relativistic velocity  $\vec{v}=c\vec{\beta}$.
The radial electric  self-field of such a rapidly moving  bunch is described by 
\cite{bbl2}
\beq
F_{\perp}(r,z)\,=\,\kappa\frac{qN\gamma}{rL}\Big\{\frac{z}{\sqrt{r^2+\gamma^2z^2}}\Big(1+ 
\frac{3}{8}\frac{b^2}{r^2}C^2_1\Big)+ 
\frac{L-z}{\sqrt{r^2+\gamma^2(l-z)^2}}\Big(1+ \frac{3}{8}\frac{b^2}{r^2}C^2_2\Big) \Big\}
\label{3.2_1}
\eeq 
with
\beq
C_1\,=\,\Big [1+\frac{\gamma^2z^2}{r^2}\Big]^{-1},\ \ \ \ 
C_2\,=\,\Big [1+\frac{\gamma^2(L-z)^2}{r^2}\Big]^{-1}.
\label{3.2_2}
\eeq 
The field  described by (\ref{3.2_1}),  has a different behavior 
at distances far apart from the bunch and in the near region, $r \le L$. At very large distances,
$r \gg \gamma z$ and $r \gg \gamma (L-z)$,  equation (\ref{3.2_1}) reduces to the Coulomb form
(\ref{3.15}). At the same time, in the near region and beyond the bunch tails, 
$\gamma z \approx \gamma (L-z)\gg r$  
and equation (\ref{3.2_1}) simplifies to
\beq
F_{\perp}\,=\,\kappa\frac{2qN}{L}\frac{1}{r}, 
\label{3-3}
\eeq 
which coincides with the external field   of a continuous beam with the linear charge
density $\lambda = qN/L$.

In an accelerator, a charged beam is influenced by its an environment (beam
pipe,  magnets, collimators, etc.), and a high-intensity
bunch induces surface (image) charges or currents into this environment. This
modifies the electric and magnetic fields around the bunch. For instance, if the bunch is 
moving in the midplane between infinitely wide  conducting planes at $x=\pm h$ 
(this simple geometry models a collimator gap), then 
the  transverse  component of the bunch electric field is described by \cite{bbl2}
\beq
F_{\perp , tot}(x)\,=\,\kappa\frac{2 qN}{Lh} \cdot\frac{\pi/2}{\sin(\frac{\pi}{2}\delta )}.
\label{3.3-23}
\eeq
where $\delta = x/h$. Thus, on the surface, $\delta =1$,  
the field is enhanced by a factor $\pi/2$    due to the presence of the conducting planes.
If the bunch is displaced  in the horizontal plane by $\bar{x}$ from the midplane 
 the resulting field behaves  as \cite{bbl2}
\beq
F_{\perp , tot}(x, \bar{x})\,=\,\kappa\frac{2 qN}{Lh} \cdot
\Big [\frac{\pi}{2}\cdot \frac{\cos(\frac{\pi}{2}\bar{\delta})}
{\sin(\frac{\pi}{2}\delta)-\sin(\frac{\pi}{2}\bar{\delta})} -\frac{\bar{\delta}}{\delta -\bar{\delta}}\Big ].
\label{3.3-24}
\eeq
Here $\bar{\delta} = \bar{x}/h$. Equations (\ref{3-3})-(\ref{3.3-24}) demonstrates that although
in the relativistic limit the electric field  does not depend on $\gamma$, the field strength may be
very high if $N$ is large and the product $Lh$ is small. 
%In the next section we analyze 

\section{\large Field Emission in HEP Colliders}

In Section 2 the conditions were described under which  FE starts to
develop. Let us examine in this section the performance of the Large Hadron Collider
(LHC) and the US Linear Collider (USLC) \cite{uslc},
 in an attempt to display their  "bottlenecks".
In the following, we shall  not  consider well established locations such as RF 
superconductive cavities, where  FE routinely  occurs and  leads to the rapid 
growth of  power dissipation  with  field. 
Our consideration here is limited to locations where bunches are moving close to conducting 
surfaces and create electric fields strong enough to activate  observable FE effects.  
For the analysis we need to know the beam parameters,  distances to the closest surfaces,
their shapes, and materials their constituent.

\begin{figure}[t]
\begin{center}
\includegraphics[height=10.0cm,width=12.0cm]{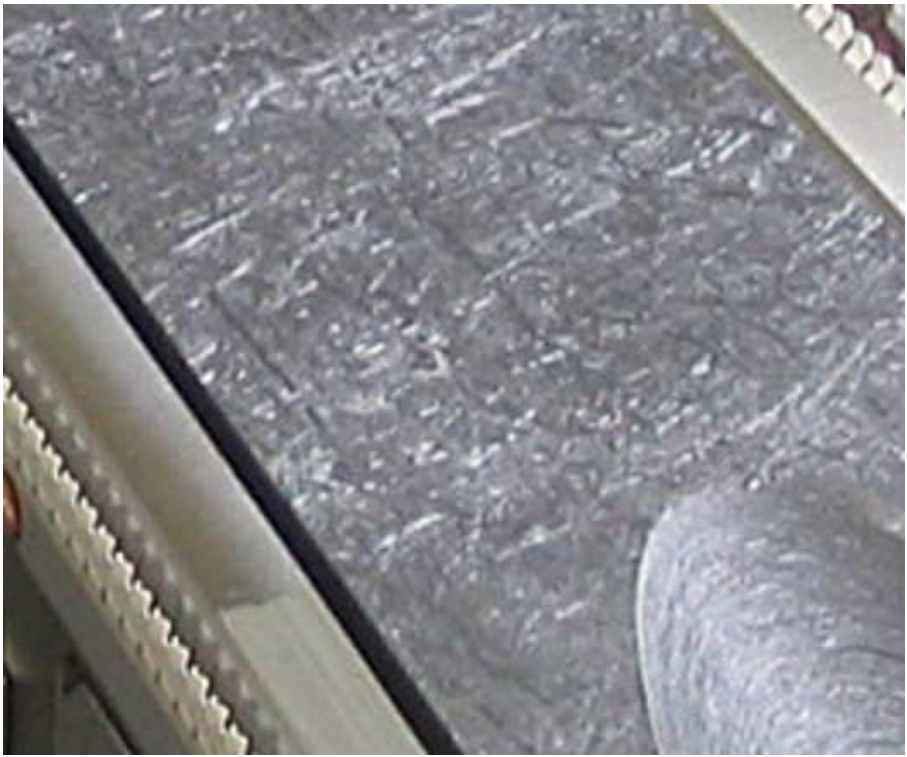}
\end{center}
\vspace*{-0mm}
\caption{ \small View from the close distance to the LHC collimator jaw
surface \cite{lhc-c01}.}
\label{fig2}
\end{figure}

\subsection{The LHC Collimator System}

Each of the two LHC rings will store 2808 bunches, each bunch populated with 
$N=1.15\cdot 10^{11}$ protons at energies of up to 7 TeV (nominal design parameters). 
To handle  the high intensity
LHC beams and the associated high loss rates of protons, a strongly acting collimator system
was designed and constructed \cite{lhc-c0}-\cite{lhc-c4}. The LHC collimators  must  restrict 
the mechanical aperture in the machine and   clean the primary halo so that quenches 
of magnets are avoided.
To provide the  required  beam quality at the interaction points,  primary and 
secondary collimators must be physically close to the high intensity beams. 
This impose strong limitations on the materials of the jaws and the length of the jaws. 
In the "final" solution for the collimation system \cite{lhc-c4}, \cite{lhc-c5},  
the  collimators in IR3 and IR7  (and IR6)
are made of graphite  with parallel jaw surfaces and small  gaps, $\sim$2 mm, at 7 TeV.
Therefore, the first candidate for a bottleneck is  the very narrow collimator gap.

Different scenarios for the beam parameters are foreseen  in the LHC operation.
  These parameters are listed in Table 1 \cite{lhc-c6}.  For the nominal scenario,
in accordance with (\ref{3.3-23}), 
the field strength at a distance of 1 mm from a bunch is equal  69 KV/cm.  
At this field level   FEC is negligible if calculated with use of the classical 
FN equation (\ref{eq:n1}). 
However,  the surface of a jaw  made of graphite may be polished only mechanically 
and the resulting surface roughness is high (see Fig \ref{fig2}). 
The  surface roughness and many other surface effects as discussed in 
Section 2  are characterized by the field enhancement parameter $\beta_{_{FN}}$, 
whose actual value needs to be determined by direct measurements.  Thereby, the  calculations 
of FEC presented below are performed for a range of $\beta_{_{FN}}$ values.
The  FEC density, $J_{_{FN}}(\beta_{_{FN}}F,T)$, calculated for the graphite jaws
 with use of  (\ref{3.3-23}) and  (\ref{s2.3.1})  at  $T$=300 K, $h$=0.1 cm and 
 $\varphi$=4.6 eV is shown in Fig. 1.  For $\beta_{_{FN}}$ below 300 and the nominal
 scenario (LHC-n), the electron emission  is  negligible. 
 FEC for $\beta_{_{FN}}$ in the range  300-500  begins to be 
 notable and at higher $\beta_{_{FN}}$  may became a very serious problem.
In the advanced scenarios (LHC-0/1), when  bunches are twice as short and their
 population is higher by the factor 1.48 (the field level on the surface  increases 
 by the factor 2.94), the emission current at $\beta_{_{FN}}$=200
 increases by 14 orders of magnitude~! At higher $\beta_{_{FN}}$, a very dense 
 electron flow will cause electrical breakdown in vacuum, jaw surface heating  
 and damage.  The electron flow could also disturb the proton beam trajectory and 
 give rise to loss of  protons. The current density depicted in Fig. 1 can be 
 translated into the number emitted electrons, $N_e$. A bunch with parameters from
 Table 1, expose the area $\delta\! A=2L\cdot\sigma^*$ during the time $\tau =L/c$ will 
 extract from  both jaw surfaces  $N_e = 2\delta\! A \tau J$  electrons.
That gives  $N_e = 7.93\cdot 10^7 J$ for  LHC-n and $N_e = 1.36\cdot 10^7 J$ for  
LHC-0/1, respectively.  Fig. 2 shows the evolution of $N_e$ with $\beta_{_{FN}}$.
We note that at  $\beta_{_{FN}}\approx$520 the number of extracted electrons $N_e$(LHC-n)
is comparable with the number of protons per bunch.

\vspace*{-2mm}

{\begin{center}
{\bf Table 1}

\vspace*{2mm}
\begin{tabular}{|r|r|r|c|c|}
\hline
\multicolumn{5}{|c|}{LHC Beam Scenarios \cite{lhc-c6,lhc-c7,lhc-c8}}\\
\hline
 bunch parameters & Nominal      & Phase 0& Phase 1& "Super-bunch"  \\
\hline 
%&&&&\\
bunch population, N         & 1.15$\times 10^{11}$   & 1.7$\times 10^{11}$ & 
1.7$\times 10^{11}$ & 5.6$\times 10^{14}$ \\
 bunch radius, $\sigma^*$     & 16.7 $\mu$m      & 11.3 $\mu$m     &   -	 & -\\ 
 bunch length, $L$   & 7.55 cm          & 3.78 cm        &    3.78 cm 	& 7500 cm\\  
 beam energy                &       7 TeV      & 7.45 TeV       &  7.45 TeV  & 7 TeV\\  
\hline
\end{tabular}
\end{center}}

\vspace*{2mm}          
\begin{figure}[t]
\hspace*{12mm}
\includegraphics[height=12.0cm,width=14.5cm]{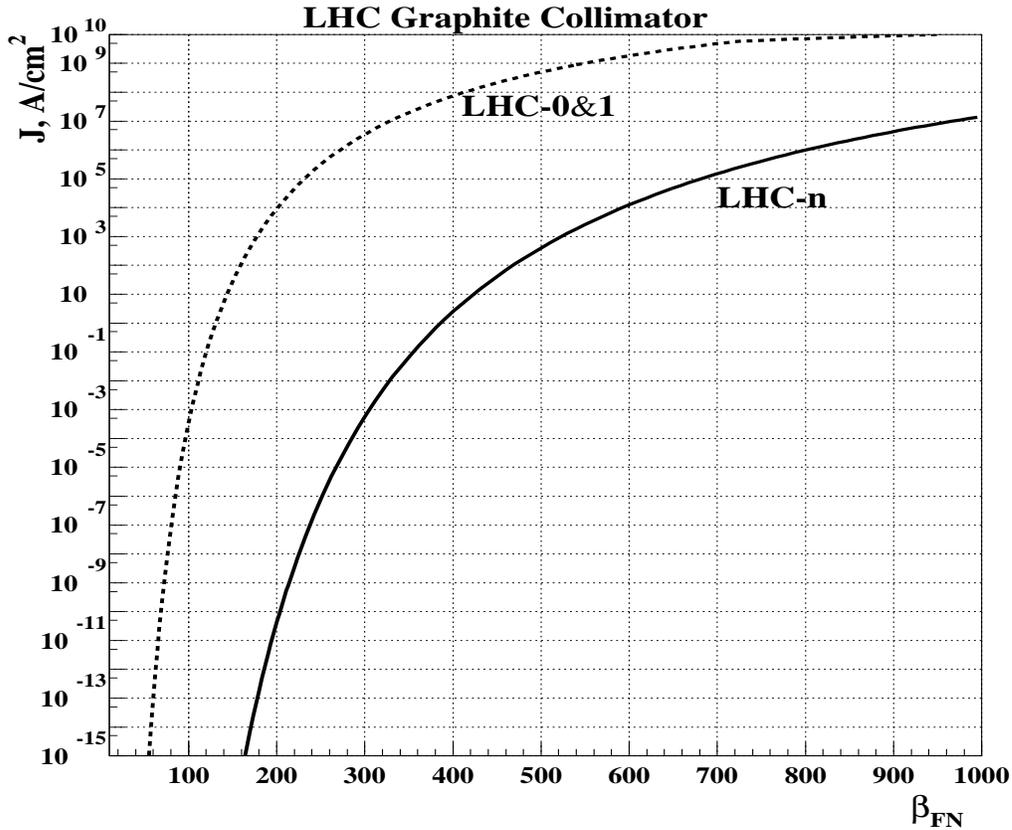}
\vspace*{-5mm}
\caption{ \small Densities of the electron current  emitted from   a graphite collimator 
jaw   at $T$=300 K,  
\newline  $\varphi$=4.6 eV   for the different beam scenarios.}
\end{figure}
%\vspace*{-10mm}
\begin{figure}[t]
\hspace*{12mm}
\includegraphics[height=12.0cm,width=14.5cm]{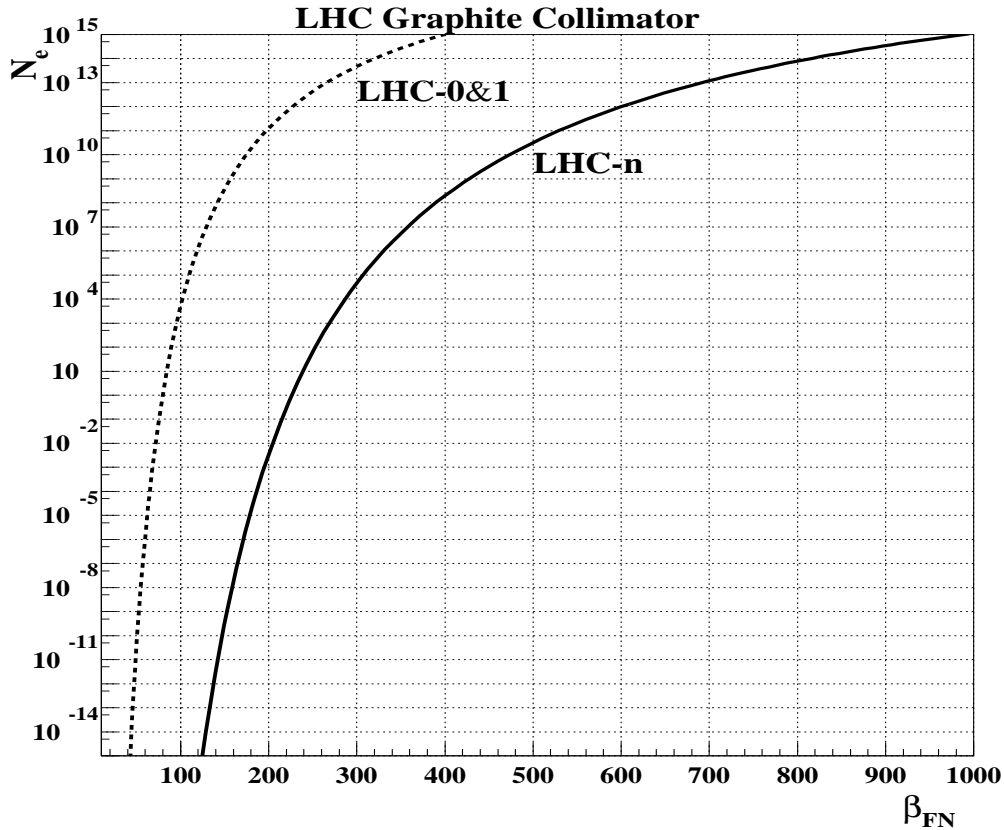}
\vspace*{-5mm}
\caption{\small  The number of electrons extracted by a single bunch from jaw surfaces 
at $T$=300 K,   
\newline   $\varphi$=4.6 eV and $h$=0.1 cm for the different beam scenarios. }
\end{figure}
\begin{figure}[t]
\hspace*{17mm}
\includegraphics[height=10.0cm,width=12.5cm]{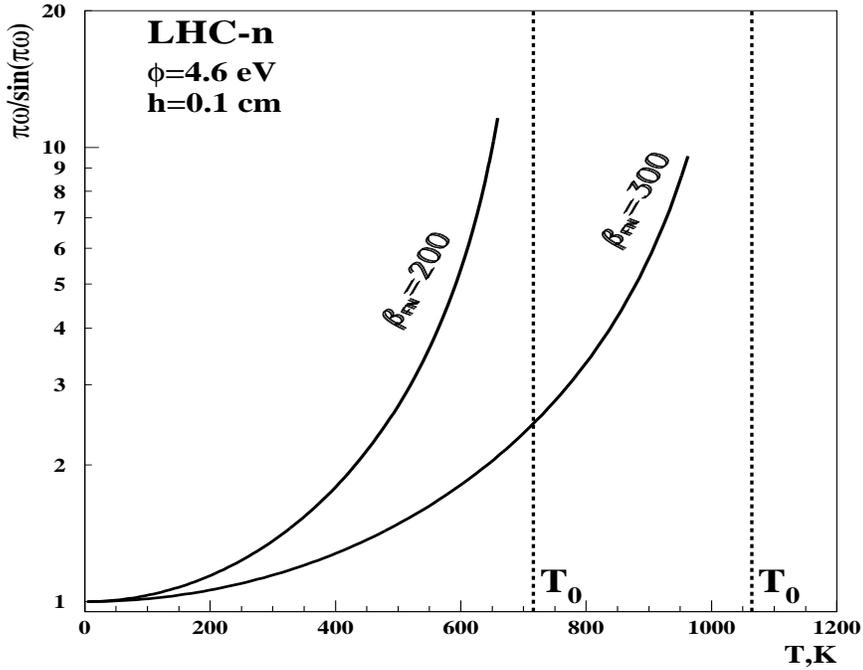}
\vspace*{-5mm}
\caption{\small The temperature  increment of the FE current (\ref{s2.3.1})
 at different $\beta_{_{FN}}$ in the temperature  \newline
 range as restricted by the conditions  (\ref{s2.g1})-(\ref{s2.g2}).  $T_0$ denotes 
 the temperature limit as determined  \newline   from  (\ref{s2.to}).  }
\end{figure}
\begin{figure}[t]
\hspace*{12mm}
\includegraphics[height=11.0cm,width=13.5cm]{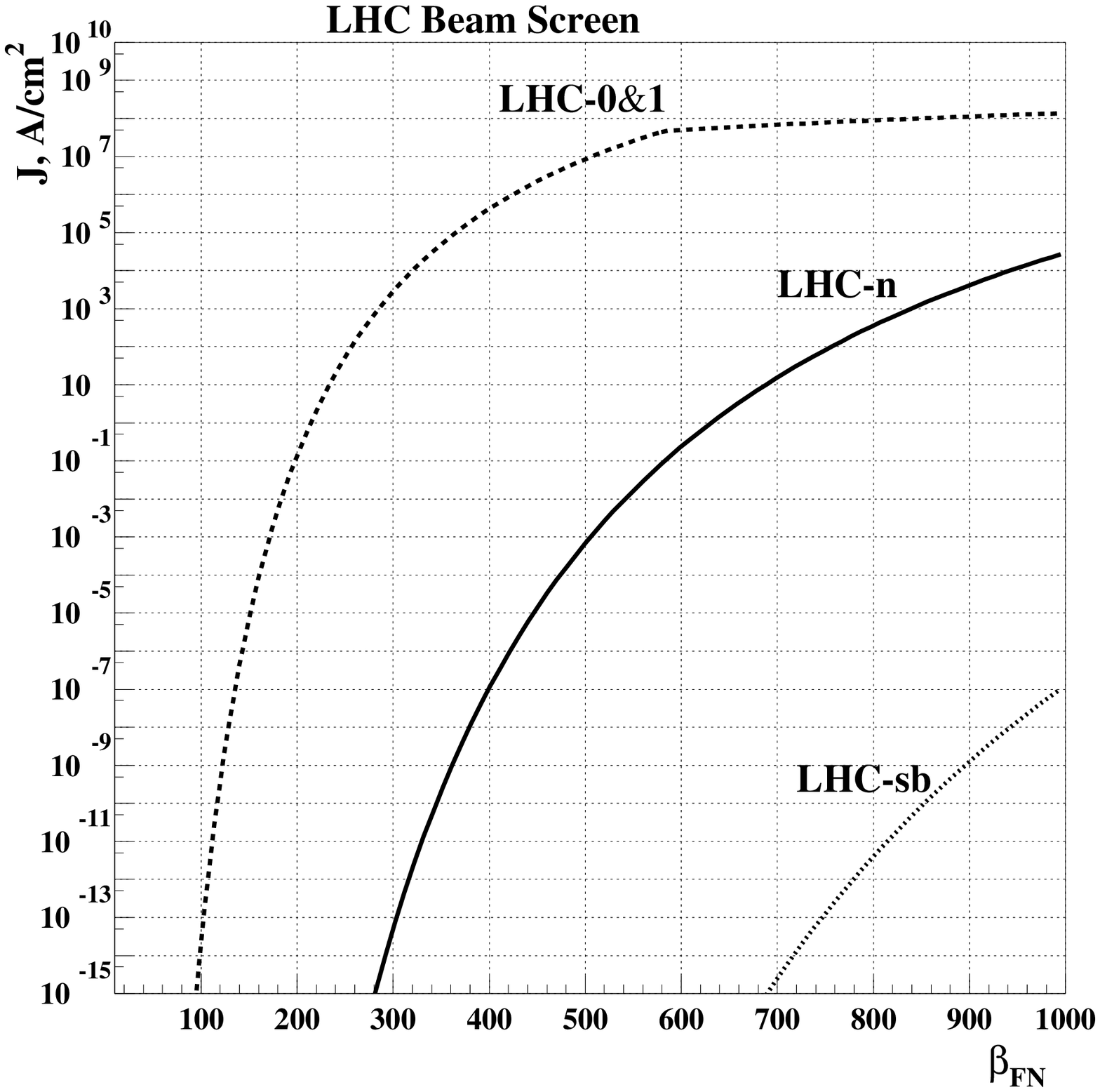}
\vspace*{-5mm}
\caption{\small   Densities of the electron current  ejected from the copper beam screen 
at $T$=20 K: \newline 
(LHC-n, LHC-0/1) due to presence of the surface   image-potential  states (IPS),  
$\varphi$(IPS)= 0.92 eV,\newline
 LHC-sb)  the super-bunch scenario , $\varphi$(Cu)= 4.65 eV. }
\label{fig4} 
\end{figure}

The FE current rises not only with $F_{eff}$, but the number of active emitters also
increases with the field. A local increase of the emission current causes  Joule heating
of the emission tip which in turn leads to  an increase of FEC. 
Fig. 3 illustrates the effect. It shows the temperature dependent factor in 
(\ref{s2.3.1}), whose rise is limited by the conditions (\ref{s2.g1}) and (\ref{s2.g2}).
For a given field strength and  higher temperatures it is necessary  to apply equations 
of the  thermo-field-emission theory \cite{MG,Chr1}. 

Joule heating by FEC is not the only mechanism responsible for an abrupt increase 
of FEC. Many studies point out that other mechanisms are also responsible for melting 
or damage of the emitter, such as the bombardment of the surface by ions. These ions 
are created by FEC as it passes through a cloud of gas evolving from the emission site
\cite{knobl-1,knobl-2}. The plasma that develops can  create an electric field on the order 
of 10$^7$ V/cm at its
interface with the conducting wall and may result in an explosion of the emitter
 \cite{LMP}-\cite{schwirz},\cite{slivkov}.

In Section 2 we noted that the image potential  states on the surface of solids are 
characterized by low values of the work function, $\varphi <$ 1 eV.  
If we  hypothesize that the field emission due to IPS contributes in an important way
 to  FEC, then the inner beam screen, consisting of
copper tubes lining the LHC vacuum chamber, is another candidate for a bottleneck. 
Indeed,  the proton beam is moving  at a distance of $\sim$2 cm  away from  
the beam screen wall with cryopumping slots \cite{lhc-01,lhc-02}, 
whose sharp edges could be very effective  emission sites with high $\beta_{_{FN}}$. 
 Fig. 4 shows the dependence of the FEC density on $\beta_{_{FN}}$ calculated with  $\varphi$=0.92 eV,
$h$=2cm and $T$=20 K, the temperature at which the beam screen will be maintained.
As seen from the figure, for the LHC-n scenario  IPS electrons will be seen 
only if $\beta_{_{FN}}>$500. However, in the LHC-0/1   scenario
emission will be large already if $\beta_{_{FN}}\sim$150 or higher.
At $\beta_{_{FN}}\approx$ 600 the top of the surface potential barrier is located below
the energy level of IPS with n=1. These electrons escape the surface freely and this is
reflected as a change in the slope of the curve. In Fig \ref{fig4} also shown calculations for the
super-bunch scenario (see Table~1) with the
nominal work function for cooper,
$\varphi$= 4.65 eV.

The results presented in Fig \ref{fig4}  are to some extent speculative.
The calculations were performed assuming that the density of IPS electrons,
pumped by the soft component of the synchrotron radiation
from the bulk to the surface 
is of the same order as the density of electrons at the Fermi level.
Nevertheless they give rise to a cause for concern
that FE from the beam screen may amplify electron cloud formation.

\subsection{The US Linear Collider}

In the similar way we may   evaluate the level of the field emission in other
colliders, for example proposed US superconducting 
linear collider (USLC) \cite{uslc} which  can became a prototype of a future
international linear collider (ILC).
 The USLC will bring into collision electrons and positrons at 
energy $\sqrt{s}$=500 GeV or, after an energy upgrade,  1000 GeV. 
The vacuum chamber in the positron ring will be coated throughout with a material 
with low secondary electron yield (e.g. conditioned titanium nitride, TiN) 
to prevent build-up of electron cloud. The beam pipe has a circular internal 
cross-section, with radius $h$=21 mm.
The vacuum chambers are all constructed from an aluminum alloy.
The bunch length  is  extraordinary  short.
The bunch compressors must reduce the $\sim$5 mm {\rm rms} length of the bunches 
extracted from the damping rings to  300 $\mu$m bunch length required 
for the main linac and  final focus systems. For optimum collider performance
in the cold option, bunches with a charge of $N=2.0\cdot 10^{10}$ and a  length 
of $L=300 \mu$m
will be injected  for  acceleration in the main linac.

\begin{figure}[t]
\hspace*{12mm}
\includegraphics[height=12.0cm,width=14.5cm]{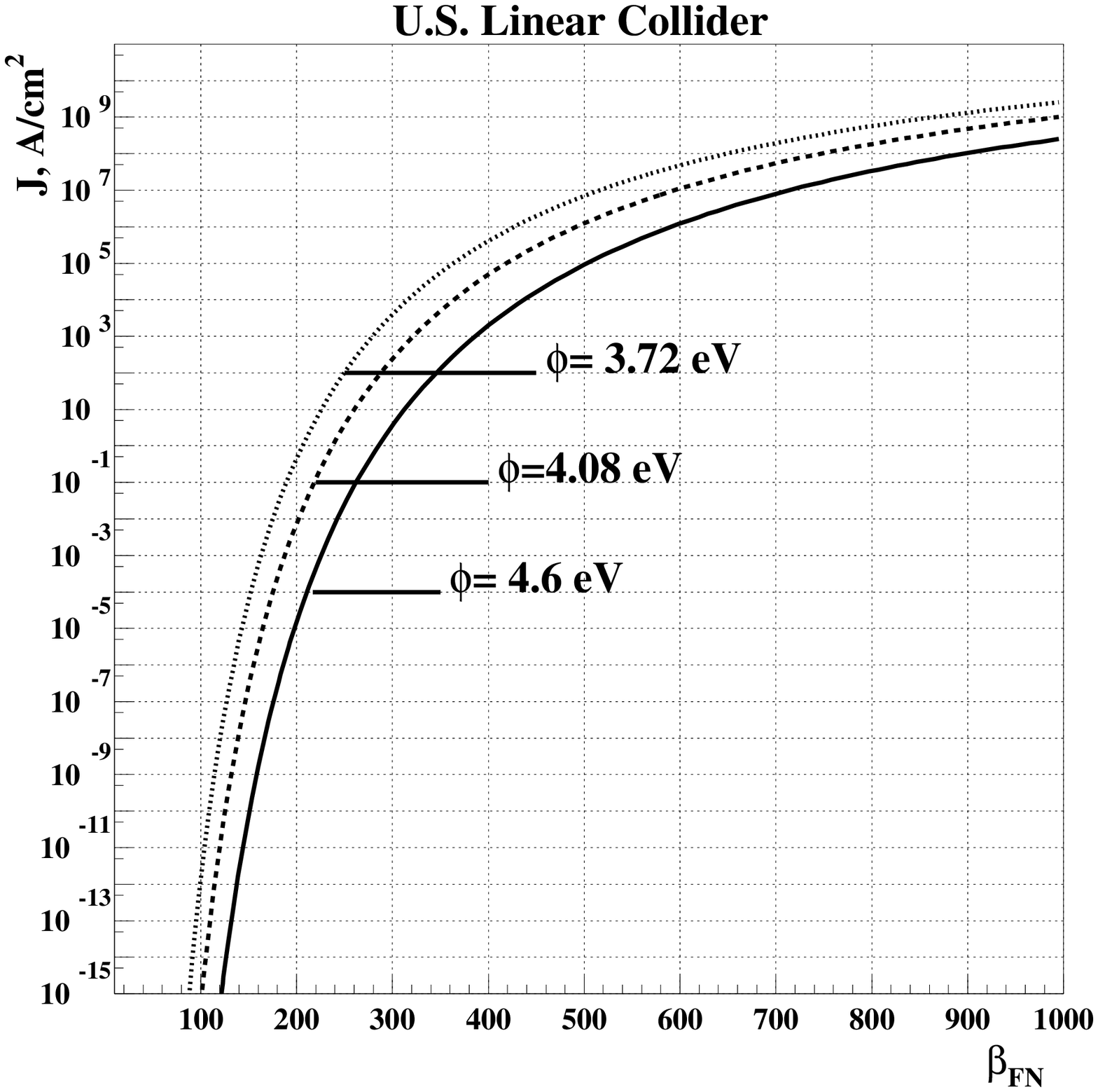}
\vspace*{-5mm}
\caption{\small Densities of the electron  current   emitted  from   the USLC main linac beam pipe   
at $T$=20 K  \newline   and various values of the work function. }
\label{fig5}
\end{figure}

For a beam well centered in the circular beam pipe, the image charge effect 
is vanished \cite{Laslett}. For this reason, instead of (\ref{3.3-23}) we use in calculations 
equation (\ref{3-3}). As a result, the  electric field  strength on an ideal smooth surface 
is characterized by the parameter $N/hL$. Direct calculations shows that 
\beq
 \Big (\frac{N}{hL}\Big )_{_{USLC}} > \frac{\pi}{2}\Big (\frac{N}{hL}\Big )_{_{LHC}},
\eeq
denoting that with the same    $\beta_{_{FN}}$ and $\varphi$ as for the LHC, 
the field emission from the USLC beam pipe surface will be stronger than in the LHC 
collimators.
Fig. \ref{fig5} shows  the FE current density predicted for a variety of  
$\varphi$ values. The range of $\varphi$  employed in the calculations reflects
 uncertainties in the work function of TiN reported by different authors 
 \cite{Fomenko}-\cite{Hanson}.
There is an observation \cite{wakab}, that changing the nitrogen concentration 
can modify the work function of TiN. Furthermore, a thickness dependency of 
the work function  was observed for TiN/Al \cite{Zheng}.
For a TiN layer with  thickness less than 20 angstroms, the TiN/Al 
gate work function is the same as reported Al work-function (about 4.08 eV).
For  TiN layers of  thickness greater than 100 angstroms, the TiN/Al gate work function 
is approximately  the same as reported TiN work-function (4.5 V). 
For  TiN layers with  thickness from 20 angstroms 
to 100 angstroms, the work function can  change as the TiN thickness is changed.
The TiN crystal orientation is important because the metal work function is strongly 
dependent on this quantity. According to \cite{yagis}, (100) orientation TiN has 
$\varphi\simeq$4.6 eV and 4.4 eV in the (111) orientation.
The work function of the TiN film was found to be sensitive to the film morphology,
stipulated by a sputtering technology and may be as low as 3.72 eV \cite{rogers}.

In this way, the uncertainty in the value of $\varphi$  gives rise to an uncertainty
in the current density of 3-5 orders of magnitude.

We finally  point out that   parameters of the positron and proton bunches
and the beam pipe  geometry at the HERA collider are such  
that even at $\beta_{_{FN}}=10^3$
the emission current density is well below $10^{-17}$ A/cm$^2$.

\begin{figure}[t]
\hspace*{12mm}
\includegraphics[height=11.0cm,width=13.0cm]{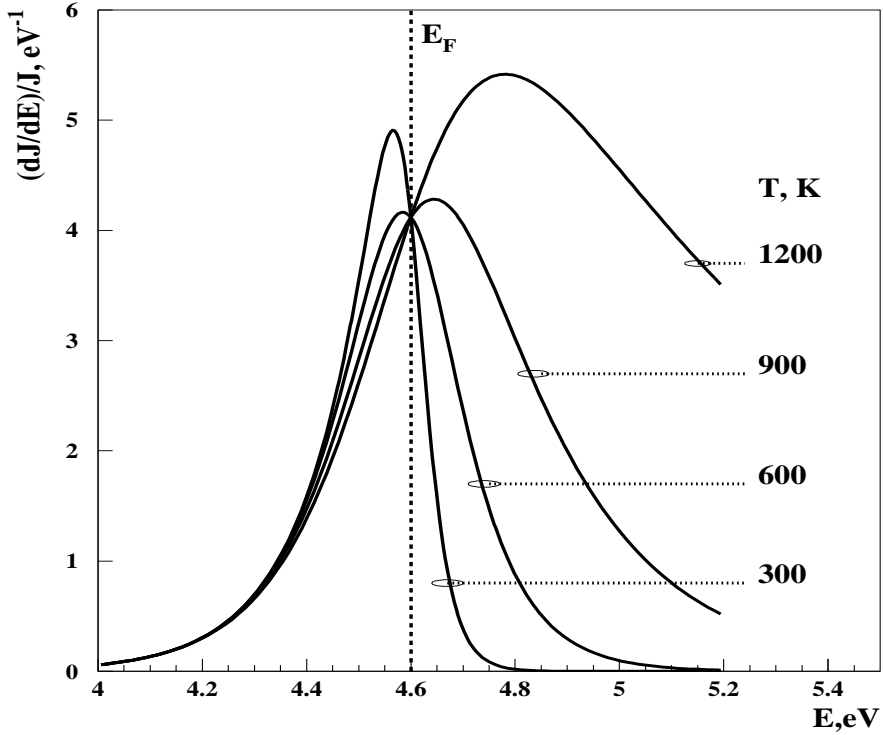}
\vspace*{-5mm}
\caption{\small  Energy distributions of the emitted electrons  for the four values of the
surface temperature. The curves are drawn according to (\ref{s2.ed}).}
\label{fig6}
\end{figure}
\begin{figure}[t]
\hspace*{12mm}
\includegraphics[height=8.0cm,width=15.0cm]{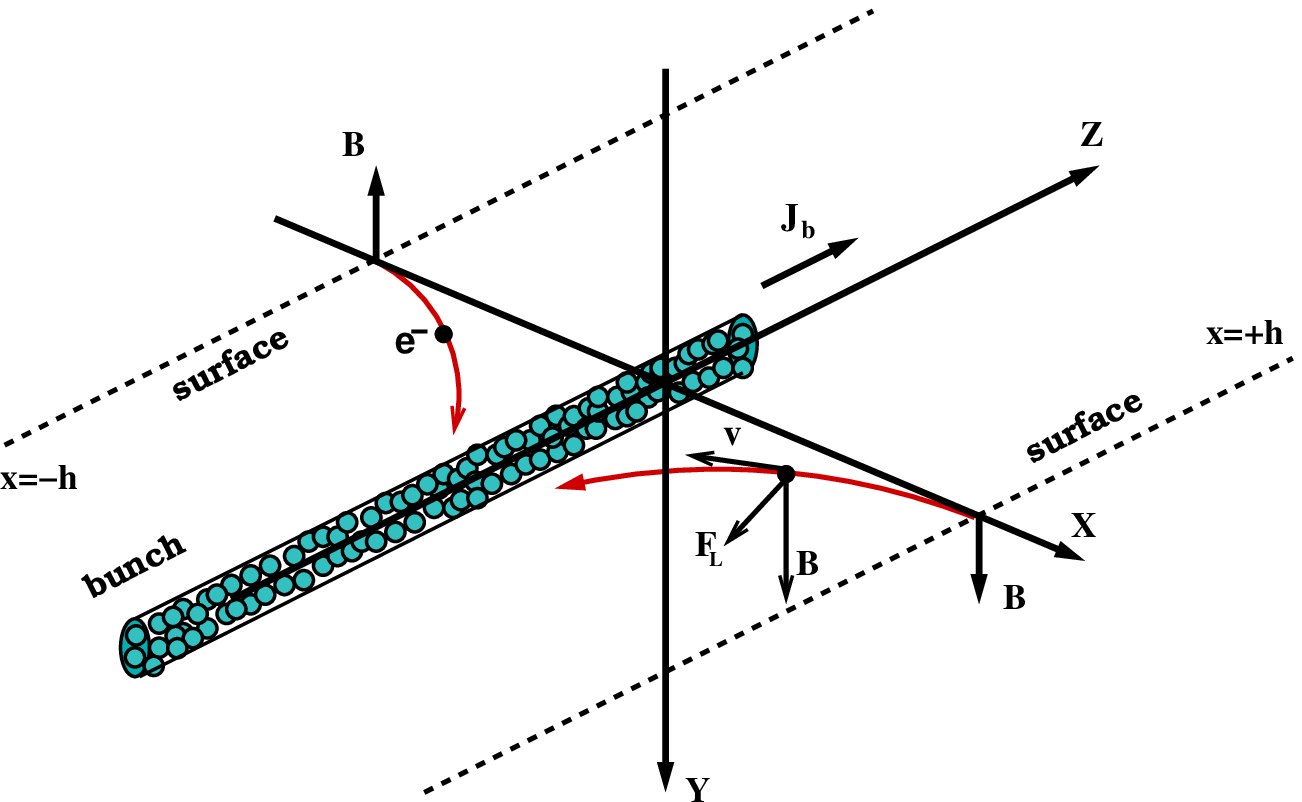}
\vspace*{+1mm}
\caption{\small Illustration to Eq. (\ref{s5-1}). Electrons ejected from the surface with the
velocity $\mathbf{v}=(\mathtt{v_x},0,0)$ are accelerated in the radial electric field  
and rotated by the  magnetic field $\mathbf{B}$, created by a cylindrical proton bunch. 
The Lorentz force acting on the electron is denoted by $\mathtt{\mathbf{F}_L}$ and 
the beam current by $\mathbf{J_b}$. }
\label{fig7}
\end{figure}

\section{\large  Electron Dynamics  in the Bunch Fields}

In this section we discuss in a semiquantitative way
the fate of electrons after their emission  in the LHC 
collimator gap.  Electrons leaves the surface with different energies.
The energy spectra of electrons  just ejected by the surface
is presented in Fig \ref{fig6}. 
For a given imposed field, the width of the energy distribution  grows
with an increase of  the surface temperature   and
more energetic electrons easier  escapes the emitter.  

Let us consider a single  electron  motion  in the self-fields of a bunch. 
The analysis is based on the classical Lorentz equation \cite{sok,land,jack}
\beq
\frac{d\vec{p}}{dt}\,=\,e\vec{F}+ec[\vec{\beta}^{(e)} \times\vec{B}].
\label{s5-1}
\eeq
%
%bremsstrahlung.
To be specific, assume that the electron is emitted from the plane surface with
the velocity vector, $\vec{\beta}^{(e)}=\vec{v}^{(e)}/c$, normal to the surface. The surface is 
placed  parallel to the ($0,\,y,\,z$) plane at a distance $x=+h$. 
The bunch is moving along the line (0,0,z) in the positive $z$ direction, as shown in 
Fig. {\ref{fig7}}.
In the simple geometry chosen, we consider, only as an illustration, the evolution 
of an electron trajectory in the ($x,\,0,\,z$) plane; in this way
 components of the self-field vectors  can be set   $\vec{F}=(F_x,0,0)$ and  
 $\vec{B}=(0,B_y,0)$, where $F_x$ is of the form (\ref{3.3-23}) and \cite{bbl2}
\beq
B_y\,=\,\frac{\kappa}{c}\frac{2qN}{Lh}\Big [ \frac{B_{_f}}{\delta} +
(1-B_{_f})\frac{\pi/2}{\sin(\frac{\pi}{2}\delta)}\Big ]\cdot \beta^{(b)}_{\!z}.
\label{s5-1.1}
\eeq
Here $\beta^{(b)}_{z}$ is the $z$-component of the bunch velocity, $\delta = x/h$, $B_{_f}=\sqrt{2\pi}L/2\pi R$ is the Laslett bunching factor and $R$ the 
average LHC machine radius. 
In the LHC-n scenario, $B_{_f}= 7.05\times 10^{-6}$ \cite{rugg,lhc-0} and 
terms proportional to  $B_{_f}$ can be omitted in  (\ref{s5-1.1}).

Surface irregularities of a very small size possess tips that  cause a local field enhancement
 and increased emission. The field of such a tip decays with distance  
approximately as \cite{Gomer}
\beq
M(\delta)=1+\beta_{_{FN}}\Big (\frac{\rho}{\rho +h|1-\delta|}\Big )^2,
\label{s5-1.2}
\eeq
where $\delta = x/h$ is a normalized distance from the emission tip, 
$\rho$ the tip radius. The self-fields $\vec{F}$ and $\vec{B}$  enter  
equation (\ref{s5-1}) multiplied by the factor (\ref{s5-1.2}).

For numerical calculations, it is convenient rewrite  (\ref{s5-1}) by 
use  the dimensionless time $\tau=t/\tau_0$ with $\tau_0=L/c$.
For a full generality,  we  retain a relativistic
treatment of equation (\ref{s5-1}) which in components becomes
\begin{eqnarray}
\frac{d(\beta^{(e)}_x \gamma^{(e)})}{d\tau}\,&=&\, (1-\beta^{(e)}_z\beta^{(b)}_{z})\cdot
\mathcal{F}(\delta) ,\nonumber\\
\frac{d(\beta^{(e)}_z \gamma^{(e)})}{d\tau}\,&=&\,-\beta^{(e)}_x\beta^{(b)}_{z}\cdot
\mathcal{F}(\delta)
\label{s5-2},
\end{eqnarray}
where the function $\mathcal{F}(\delta)$ outside the bunch takes the form
\beq
\mathcal{F}(\delta)=\mathcal{A}\cdot\frac{M(\delta)}{\sin(\frac{\pi}{2}\delta )},
\ \ \ \ \ \sigma^*<|x|\leq h,
\label{s5-21}
\eeq
and inside the bunch is \cite{bbl2}
\beq
\mathcal{F}(\delta)=\frac{2\mathcal{A}}{\pi}\Big 
(\frac{h^2}{{\sigma^*}^2} +\frac{\pi^2}{24}\Big )\delta,\ \ \ \ \ |x|\leq\sigma^*
\label{s5-22}
\eeq
with the dimensionless constant 
$$\mathcal{A}=\frac{\kappa\pi}{mc}\frac{ eqN}{Lh}\tau_0.$$
A direct estimation gives that $\mathcal{A}=O(1)$ for  the LHC-n parameters.
Fig \ref{fig7} schematically shows  electron trajectories and forces acting 
on  the electron as a consequence of  equation (\ref{s5-1}).

Let us analyze the system (\ref{s5-2}). 
For a relativistic bunch one may put 
$\beta^{(b)}_{z}=+1$  and therefore below we omit  superscripts in the electron 
velocity components.
A partial solution of  (\ref{s5-2}) can be found for  nonrelativistic electrons,
 $\gamma\simeq 1$, 
by taking a ration of the two equations. That gives
$$\beta_x\frac{d\beta_x}{d\tau} = (\beta_z-1)\frac{d\beta_z }{d\tau}.$$
After integration of the last equation and some algebra one finds a solution  
in the form
$$ \beta_z = 1-\sqrt{1+\beta^2_x-\beta^2_0}\,, $$
where the integration constant $\beta_0$ is fixed by  initial conditions $\beta_z =$0,  
$\beta_x=\beta_0$ at $t$=0. This equation connects the  $p_z$ and $p_x$ 
components of the electron momentum in the following way
\beq
p_z= - \frac{{p^2}_{\!\!\! _{x}}-{p^2}_{\!\!\! _{0}}}{2mc}.
\label{s5-3.1} 
\eeq
Equation (\ref{s5-3.1}) implies that during  the electron acceleration  
$p_z$ remains negative and  much smaller than $p_x$. Thus, the electron displacement
in the $z$ direction is expected to be small too.
 
The magnetic field of the bunch gives rise to the $p_z$ component 
 and  the electron  turn in the ($x$,0,$z$) plane.     For an accelerated charge
 the radiation emitted at any instant is approximately the same as that emitted by a
 particle moving instantaneously along the arc of a circular path whose radius of
 curvature is $R_i$. The frequency spectrum of radiation emitted by a charge in
 instantaneously circular motion is characterized by the critical frequency
\beq 
\omega_c = \frac{3}{2}\frac{c}{R_i}\gamma^3,
\label{s5-4} 
\eeq
beyond which there is negligible radiation at any angle \cite{sok,jack}. The corresponding 
energy of the photon
is $E_{_\gamma}= \hbar\omega_c$.

For a nonrelativistic charge whose motion is  described by the system (\ref{s5-2})
with $\gamma^{(e)}=1$,  
 the instantaneous radius $R_i$ can be
found in the following way. Let  the point ($x_0$, $z_0$) be the center of
the circle of   radius $R_i$ 
\beq 
(x-x_0)^2+(z-z_0)^2={R^2}_{\!\! i}
\label{s5-5} 
\eeq
whose infinitesimal arc with the point ($x$, $z$) on it coincide at the  moment $\tau$ with the electron 
trajectory. To find $R_i$ we differentiate (\ref{s5-5}) twice by 
 $\tau$  and find relations between $x-x_0$, $z-z_0$ and $\beta_x$, $\beta_z$, 
 $\dot{\beta_x}$ and  $\dot{\beta_z}$. Using  (\ref{s5-2}) for $\dot{\beta_x}$ and $\dot{\beta_z}$, 
 we obtain 
\beq
R_i(\tau)=\frac{c\tau_0\beta^3}{|\mathcal{F}(\delta)|(\beta^2_x - \beta^2_z +\beta_z)}.
\label{s5-6} 
\eeq
%  
%The total power $W$ radiated off per unit time is described by the Larmor
%formula which for the  case under consideration transformed into
%
%\beq
%P=\frac{dW}{d\tau}=\frac{2}{3}\frac{\kappa e^2}{L}\Big (\dot{\beta}\Big )^2
%\eeq
%
%where $\dot{\beta}$ can be evaluated with use of (\ref{s5-2}).

Now we are in a position to present  numerical results. The input
parameters for a numerical solution of the system (\ref{s5-2}) are listed in Table 2. These are
the distance, $h$, from the bunch center to the collimator plane, the tip radius $\rho$ in equation 
(\ref{s5-1.2}),
the field enhancement factor at the surface, $\beta_{_{FN}}$, the electron velocity $\beta$, 
or its kinetic 
energy $E_{_{kin}}$. Parameters of the proton beam are correspond to the LHC-n scenario, 
see Table 1.

\vspace*{4mm}
{\begin{center}
{\bf Table 2}
\vspace*{2mm}

\hspace*{-4mm}
\begin{tabular}{|c|c|c|c|c|c|}
\hline
\multicolumn{6}{|c|}{Input  paramerets for the system (\ref{s5-2})  }\\
\hline
$h$, mm & $\rho$, m        & $\beta_{_{FN}}$ & $E_{_{kin}}$, eV& $\beta_x$& $\beta_z$\\
\hline 
1 & 1.0$\times 10^{-8}$& 300             &     4.6       &   -4.2$\times 10^{-3}$ & 0.0\\                                           
\hline
\end{tabular}
\end{center}}
\vspace*{4mm}

Table 3  presents  parameters of the electron trajectory,
 whose numerical values were calculated
 at the emission  point ($\delta$=1), at the  bunch center ($\delta$=0) 
and at the opposite collimator plate ($\delta =-$1).
The trajectory is split into three parts. 
At the first stage,  the electron is accelerating 
by the  field (\ref{s5-21}) and moves toward the bunch (Fig \ref{fig7}). 
The local field enhancement (\ref{s5-1.2}) causes an extremely high acceleration rate and a rapid increase 
of the electron momentum. At the start time the instant curvature of the trajectory $R_i$ is very small and 
 the energy loss  $E_{_\gamma}$ through radiation can  run up to 5\% of the initial electron energy.    
At later times with an increase  in the distance from  the surface, the energy loss by radiation 
is very low, $E_{_\gamma}\sim 10^{-6}-10^{-4}$ eV.

At the second stage, the electron is crossing the bunch.
The internal field of the bunch is described by (\ref{s5-22}).
At the bunch center
the electron momentum  reaches a maximum value of 151 KeV/c  
 at $\tau=$0.129 after emission. An average distance between protons in the LHC-n 
 bunch is of the order of 100 nm and the electron   either passes through the bunch  or will

\vspace*{3mm}
%\newpage
{\begin{center}
{\bf Table 3}
\vspace*{2mm}

\hspace*{-4mm}
\begin{tabular}{|l|r|c|c|c|c|c|c|c|}
\hline
\multicolumn{9}{|c|}{Selected paramerets of the electron trajectory }\\
\hline
$\delta$& $\tau$ & $\beta$     &$E_{_{kin}}$, eV   & p, KeV/c& $R_i$, m     & $\omega_c$, Hz   &$E_{_\gamma}$, eV
& $\sqrt{s_{ep}}$, GeV \\
\hline 
1 & 0    & 4.2$\times 10^{-3}$& 4.6               & 2.17    & 1.1$\times 10^{-6}$& 4.3$\times 10^{14}$&0.283 & - \\                                           
0 & 0.129& 0.293              & 2.3$\times 10^{4}$& 151.0   & 7.7$\times 10^{-2}$& 6.7$\times 10^{9}$&4.4$\times 10^{-6}$ & 2.94 \\ 
-1& 0.254& 3.3$\times 10^{-2}$& 271.1             & 16.6    & 4.8$\times 10^{-3}$& 9.5$\times 10^{10}$&6.2$\times 10^{-5}$ & -\\
\hline
\end{tabular}
\end{center}}

\vspace*{4mm}

\noindent
   be scattered by a proton.    
In the later case, the electron and the proton undergo an elastic or inelastic collision 
at the center of mass energy of 2.94 GeV. In the rest frame of the proton, the electron
carries the  momentum of 4.13 GeV/c. Such an amount of energy is enough for hadron production
in an inelastic $ep$ collision.

At the last stage, the electron is  decelerated   by the bunch field.  
If the electron  passes the bunch without scattering, then
at the moment when $\tau=$0.254 it impacts  the opposite  plate of the collimator at $\delta =$-1
with a kinetic energy of 271 eV. For a fully symmetric
configuration and neglecting the electron energy loss  in the course of   acceleration and
deceleration, the electron has to arrive at the opposite plate with the same kinetic energy
as at the start point. The energy difference arise  due to an additional  
acceleration of the electron
 in the enhanced field at the tip apex and the assumption that at 
the end point the surface is flat.

Above, we considered the emission of a single electron, 
when at   $\tau$=0  the  emission  point and the bunch head 
 had the same z-coordinate.
However, the field  emission   occurs throughout the entire region $L$, 
covered by the field of the bunch. 
Each electron reaches the bunch center after time $\Delta\tau\approx$ 0.129.
If  now account a bunch movement, then electrons
 emitted at $1-2\Delta\tau < \tau < 1$, 
will experience only initial acceleration or deceleration in the wake field of the bunch.
Thus, the   electron energy spectrum  will be very broad, from 4.6 eV to 23 KeV.
As a result of  emission  from   both collimator plates, the  proton bunch occurs in 
the electron streams being intersected.
For high $\beta_{_{FN}}$ it is  necessary  take into account the  space charge
of  electron clouds, whose electric field reduces the the strength of the bunch field
 and leads to a decrease of the electron emission. 
Nevertheless, the leading part of the bunch of  length $\Delta\tau$ will
 always remains out of the electron cloud and  triggers the field emission.

%However, after being displaced further, proton bunch is freed from the  electron cloud, 
% the field strength is restored and emission again grows.
%As a result of this process the emission of electrons along the  bunch path
% is characterized by  specific  space pulsations with the frequency  $\omega \sim$ c/L.

The interaction between the electron and the surface results in the ejection of 
secondary electrons from the material. 
The secondary electrons consist of true  secondaries and those elastically reflected.
The number of secondary electrons is given by the secondary emission
coefficient (SEC) that depends on the surface characteristics and 
on the impact energy of the primary. In turn, the secondaries are 
accelerated and, on impact, produce further
generation of electrons.   The electron-cloud build-up is sensitive 
to the  intensity, spacing, and length of the proton bunches. 
The electron flow increases exponentially if the number of  emitted 
electrons exceeds the number of impacting electrons, and
if their trajectories satisfy some specific conditions. 
For most materials the SEC exceeds unity  for impact energies 
in a range from a few tens of eV to a few thousand of eV. This type of 
electron multiplication, so-called multipacting, and build-up  a cloud 
of electrons has been studied recently very intensively \cite{lhc-01}-\cite{lhc-10}. 
In the LHC arcs, as believed,  the dominant source of electrons 
will be photo-electrons from  synchrotron radiation and beam-induced 
multipacting  to be the leading source of sustained electron-production.

The region near the scrapers and collimators is susceptible 
to a high beam-loss, and is potentially another 
location of high electron concentration. Protons incident on
the collimator surfaces produce secondary electrons. Depending 
on the energy of the beam and the incident angle,
the secondary electron-to-proton yield can greatly exceed unity
 when the incident beam is nearly parallel to the surface.

In  conclusion let us note that any heterogeneous structures
such as  a grooved metal surface \cite{ilc-1,ilc-2} introduced with intention to combat 
with the multiplication of electrons
and located near an intensive positively charged  beam,  may 
in  real conditions actually provoke  field emission.

\section{\large \bf {\it Praemonitus Praemunitus }}
%Forewarned, forearmed.
%
 The analysis performed in the previous sections makes it possible to draw 
the conclusion that in the addition to the known electron sources, 
 electron field emission  intensified by multipacting 
can make a dominant contribution to the build-up of an  electron 
cloud in the LHC collimator system and  may become a serious problem.
The  analysis of the ILC  prototype reveals that a noticeable field emission will 
accompany positron bunches on their entire path during acceleration.

From the examples  considered we learn that the level of  field emission  is controlled  
by two essential parameters. The first of these, $\beta_{_{FN}}$, is wholly determined 
by the state of the emitting surface and by the processes proceeding on it. 
Therefore, control of $\beta_{_{FN}}$ is possible only within  certain limits.
The effect of surface aging  will increase $\beta_{_{FN}}$ with time.
The actual value of $\beta_{_{FN}}$ for the graphite jaw can be obtained
only by direct measurement.

For $\beta_{_{FN}}$ below 300 and the nominal  LHC-n
 scenario, the electron emission  is  negligible.  
 If the $\beta_{_{FN}}$ value is in the range  of 300-500, FEC  begins to be 
 notable. At higher  $\beta_{_{FN}}$, FEC  may became a very serious problem.
In the advanced scenarios (LHC-0/1), when  bunches are twice as short and their
 population is higher by a factor 1.48,  the emission current at $\beta_{_{FN}}$=200
 increases by 14 orders of magnitude as compared with LHC-n. At higher $\beta_{_{FN}}$, a very dense 
 electron flow will cause electrical breakdown in vacuum, jaw surface heating  
 and damage.  The electron flow could also disturb the proton beam trajectory and 
 give rise to a loss of  protons.
 
The value of the  second  parameter, N/hL, is assigned by the beam parameters and 
the distance to the conducting surface. It is possible and necessary to choose 
the components of $\beta_{_{FN}}$N/hL in the way to minimize the electron field emission.

\vspace*{4mm}
\noindent
{\it \bf Acknowledgments.} The author is grateful to P. Bussey, P.F.~Ermolov,
E. Lohrmann, E.B.~Oborneva, V.I.~Shvedunov and A.N.~Skrinsky
for   useful discussions and comments. A special thanks goes to O. Br\"{u}ning whose
talk \cite{lhc-c6} stimulated the interest of the author to problems discussed 
in the paper.

%%%%%%%%%%%%%%%%%%%%%%%%%%%%%%%%%%%%%%%%%%%%%
\newpage
\appendix
\renewcommand{\theequation}{A-\arabic{equation}}
\setcounter{equation}{0}  % reset counter
\section*{ \large Appendix: 
{\normalsize Electric Field of a Relativistic Dipole}
} % use *-form to suppress numbering
\vspace*{4mm}

Take two point charges, +q and -q, separated by a distance $2h$. 
The charges are at rest in  frame ${\bf \bar{S}}$ and the $\bar{x}$-axis passes through 
the charges. 
We denote the coordinates of a point in the frame ${\bf \bar{S}}$ by $(\bar{x},\bar{y},\bar{z})$.
Then the potential from the two  charges (a dipole) is given by

\beq
\varphi (\bar{x},\bar{y},\bar{z})\,=\,\kappa\Big\{ \frac{q}{\sqrt{(\bar{x}-h)^2+\bar{y}^2+\bar{z}^2}}
- \frac{q}{\sqrt{(\bar{x}+h)^2+\bar{y}^2+\bar{z}^2}}
\Big\}.
\label{a1}
\eeq
We obtain the  electric field of a dipole at rest
 by the usual rule:
\beq
\vec{F}\,=\,-\,\vec{\nabla}\varphi
\label{a2}
\eeq
or in components
\beq
\bar{F}_x= \kappa q \Big [\frac{\bar{x}-h}{\bar{D}^3_{-}}\,-\,\frac{\bar{x}+h}{\bar{D}^3_{+}}\Big ],
\ \ \
\bar{F}_y= \kappa q \Big [\frac{1}{\bar{D}^3_{-}}\,-\,\frac{1}{\bar{D}^3_{+}}\Big ]\,\bar{y}, 
\ \ \
\bar{F}_z= \kappa q \Big [\frac{1}{\bar{D}^3_{-}}\,-\,\frac{1}{\bar{D}^3_{+}}\Big ]\,\bar{z},
\label{a3}
\eeq
where 
\beq
\bar{D}_{\pm}\,=\,\sqrt{(\bar{x}\pm h)^2+\bar{y}^2+\bar{z}^2}.
\label{a4}
\eeq
Now, suppose that the dipole is moving along the z-axis of the frame ${\bf S}$ with 
 velocity $v$.
Then the space coordinates in ${\bf S}$ and ${\bf \bar{S}}$ are related by the Lorentz 
equations
\beq
\bar{x}\,=\,x,\ \ \ \bar{y}\,=\,y,\ \ \ \bar{z}\,=\,\gamma (z-vt), 
\label{a6}
\eeq
The transformation laws for the components of the electric 
field
% $F$  and magnetic $H$ fields 
can be written as
\beq
F_x\,=\,\gamma \bar{F}_x,\ \ \ F_y\,=\,\gamma \bar{F}_y,\ \ \  F_z\,=\, \bar{F}_z.
\label{a7}
\eeq
Substituting expressions of $\bar{F}$ from (\ref{a3}) into (\ref{a7}) and  account  
(\ref{a6}), we arrive at
\begin{eqnarray}
F_x&=& \kappa q \gamma\Big [\frac{x-h}{D^3_{-}}\,-\,\frac{x+h}{D^3_{+}}\Big ]\\
F_y&=& \kappa q \gamma\Big [\frac{1}{D^3_{-}}\,-\,\frac{1}{D^3_{+}}\Big ]\,y \\
F_z&=& \kappa q \gamma\Big [\frac{1}{D^3_{-}}\,-\,\frac{1}{D^3_{+}}\Big ]\,(z-vt),
\label{a8}
\end{eqnarray}
where
\beq
D_{\pm}\,=\,\sqrt{(x\pm h)^2+y^2+\gamma^2(z-vt)^2}.
\label{a7*}
\eeq
On the plane $x=0$
\beq
F_x\,=\,-\,\frac{2\kappa q\gamma h}{\big [ h^2+y^2+\gamma^2(z-vt)^2\, \big ]^{3/2}},\ \ \
F_y\,=\,F_z\,=\,0.
\label{a9}
\eeq

{}
\end{document}